\documentstyle[12pt,graphicx]{article}
\begin{document}
\baselineskip 0.6cm

\def\simgt{\mathrel{\lower2.5pt\vbox{\lineskip=0pt\baselineskip=0pt
           \hbox{$>$}\hbox{$\sim$}}}}
\def\simlt{\mathrel{\lower2.5pt\vbox{\lineskip=0pt\baselineskip=0pt
           \hbox{$<$}\hbox{$\sim$}}}}

\begin{titlepage}

\begin{flushright}
SLAC-PUB-11462 \\
UCB-PTH-05/24 \\
LBNL-58773
\end{flushright}

\vskip 1.2cm

\begin{center}

{\Large \bf 
A Solution to the Supersymmetric Fine-Tuning Problem
\\
within the MSSM
}

\vskip 1.0cm

{\large
Ryuichiro Kitano$^a$ and Yasunori Nomura$^{b,c}$}

\vskip 0.4cm

$^a$ {\it Stanford Linear Accelerator Center, 
                Stanford University, Stanford, CA 94309} \\
$^b$ {\it Department of Physics, University of California,
                Berkeley, CA 94720} \\
$^c$ {\it Theoretical Physics Group, Lawrence Berkeley National Laboratory,
                Berkeley, CA 94720} \\

\vskip 1.2cm

\abstract{Weak scale supersymmetry has a generic problem of fine-tuning 
in reproducing the correct scale for electroweak symmetry breaking.  The 
problem is particularly severe in the minimal supersymmetric extension 
of the standard model (MSSM).  We present a solution to this problem that 
does not require an extension of the MSSM at the weak scale.  Superparticle 
masses are generated by a comparable mixture of moduli and anomaly mediated 
contributions, and the messenger scale of supersymmetry breaking is 
effectively lowered to the TeV region.  Crucial elements for the solution 
are a large $A$ term for the top squarks and a small $B$ term for the 
Higgs doublets.  Requiring no fine-tuning worse than $20\%$, we obtain 
rather sharp predictions on the spectrum.  The gaugino masses are almost 
universal at the weak scale with the mass between $450$ and $900~{\rm GeV}$. 
The squark and slepton masses are also nearly universal at the weak scale 
with the mass a factor of $\sqrt{2}$ smaller than that of the gauginos. 
The only exception is the top squarks whose masses split from the other 
squark masses by about $m_t/\sqrt{2}$.  The lightest Higgs boson mass is 
smaller than $120~{\rm GeV}$, while the ratio of the vacuum expectation 
values for the two Higgs doublets, $\tan\beta$, is larger than about $5$. 
The lightest superparticle is the neutral Higgsino of the mass below 
$190~{\rm GeV}$, which can be dark matter of the universe.  The mass of 
the lighter top squark can be smaller than $300~{\rm GeV}$, which may 
be relevant for Run~II at the Tevatron.}

\end{center}
\end{titlepage}

\section{Introduction and Summary}
\label{sec:intro}

One of the primary reasons for looking for physics beyond the standard 
model comes from the concept of naturalness.  In the standard model, the 
Higgs mass-squared parameter receives radiative corrections of order the 
cutoff scale squared, so unless there is an unnatural cancellation we 
expect that the cutoff of the standard model is not much larger than the 
scale of electroweak symmetry breaking.  Weak scale supersymmetry provides 
an excellent candidate for physics above this scale, in which the role 
of the cutoff for the Higgs mass-squared parameter is played essentially 
by superparticle masses.  The theory can then be extrapolated up to 
very high energies, giving the successful prediction for gauge coupling 
unification~\cite{Dimopoulos:1981zb}.

The negative result for the Higgs boson search at the LEP~II, however, 
brings some doubt on this picture, at least the one based on the minimal 
supersymmetric extension of the standard model (MSSM).  In the MSSM, the 
mass of the physical Higgs boson, $M_{\rm Higgs}$, is smaller than the 
$Z$ boson mass at tree level.  The mass can be pushed up to larger values 
by a top-stop loop contribution~\cite{Okada:1990vk}, but the LEP~II 
bound of $M_{\rm Higgs} \simgt 114.4~{\rm GeV}$~\cite{Barate:2003sz} 
requires this contribution to be rather large.  For a reasonably 
small value for the stop mixing parameter, such a large contribution 
arises only for rather large top squark masses, $m_{\tilde{t}} \simgt 
(800\!\sim\!1200)~{\rm GeV}$~\cite{Carena:1995wu}.  This is a problem, 
because the (up-type) Higgs boson mass-squared parameter, $m_{H_u}^2$, 
then receives a large contribution
\begin{equation}
  \delta m_{H_u}^2 \simeq -\frac{3y_t^2}{4\pi^2}\, m_{\tilde{t}}^2\, 
    \ln\Biggl( \frac{M_{\rm mess}}{m_{\tilde{t}}} \Biggr),
\label{eq:corr-Higgs}
\end{equation}
where $M_{\rm mess}$ is the scale at which the squark and slepton masses 
are generated.  For $M_{\rm mess}$ of order the unification scale, for 
example, this gives $|\delta m_{H_u}^2|^{1/2}$ larger than a TeV.  Since 
the size of the Higgs quartic coupling is essentially determined by 
supersymmetry, we then find that this contribution must be canceled with 
some other contribution to $m_{H_u}^2$ at a level of one percent or even 
worse, to reproduce the correct scale for electroweak symmetry breaking, 
$v \simeq 174~{\rm GeV}$.  In fact, the problem becomes even keener if 
the top quark is relatively light as suggested by the latest experimental 
data: $m_t = 172.7 \pm 2.9~{\rm GeV}$~\cite{Group:2005cc}.  Since the 
correction to the Higgs boson mass is roughly proportional to $m_t^4 
\ln(m_{\tilde{t}}/m_t)$, such a light top quark pushes up the lowest 
value of $m_{\tilde{t}}$ giving $M_{\rm Higgs} \simgt 114.4~{\rm GeV}$ 
quite high, e.g. $m_{\tilde{t}} \simgt 1~{\rm TeV}$. 

A simple way to avoid the above problem, called the supersymmetric 
fine-tuning problem, is to make both $m_{\tilde{t}}^2$ and $\ln(M_{\rm 
mess}/m_{\tilde{t}})$ small.  For $\ln(M_{\rm mess}/m_{\tilde{t}})$ 
as small as a factor of a few, the fine-tuning can be avoided for 
$m_{\tilde{t}} \simlt (400\!\sim\!510)(M_{\rm Higgs}/150~{\rm GeV})
(3.5/\ln(M_{\rm mess}/m_{\tilde{t}}))^{1/2}$~\cite{Chacko:2005ra,%
Nomura:2005qg}, where the numbers correspond to making fine-tuning 
better than $(20\!\sim\!30)\%$.  Such light top squarks can easily be 
accommodated if we introduce an additional contribution to the Higgs 
boson mass.  (For theories giving such a contribution, see e.g.~%
[\ref{Ellis:1988er:X}~--~\ref{Babu:2004xg:X},~\ref{Chacko:2005ra:X}].) 
This, however, necessarily requires a deviation from the MSSM at the 
weak scale. 

In this paper we study the question: is it possible to solve the 
supersymmetric fine-tuning problem without extending the MSSM at 
the weak scale?  The difficulty of doing that is numerically clear.  
The amount of fine-tuning, $\Delta^{-1}$, is measured by the ratio of 
the left-hand-side to the largest contribution to the right-hand-side 
in the equation determining the weak scale, $M_{\rm Higgs}^2/2 \simeq 
-m_{H_u}^2 - |\mu|^2$~\cite{Chacko:2005ra-2}.  This gives
\begin{equation}
  \Delta^{-1} \approx \frac{M_{\rm Higgs}^2}{2\, \delta m_{H_u}^2},
\label{eq:fine-tuning}
\end{equation}
where $\delta m_{H_u}^2$ is the top-stop contribution of 
Eq.~(\ref{eq:corr-Higgs}).  Suppose now that the superparticle masses 
are generated at one loop through standard model gauge interactions. 
In this case, $M_{\rm mess}$ can be as small as $(12\pi^2/g_3^2)^{1/2} 
m_{\tilde{t}} \simeq 8~{\rm TeV}$, where $g_3$ is the $SU(3)_C$ gauge 
coupling at $M_{\rm mess}$, so that $\ln(M_{\rm mess}/m_{\tilde{t}})$ 
can be as small as $\simeq 2.3$.  Even with this small logarithm, 
Eq.~(\ref{eq:fine-tuning}) still gives fine-tuning as bad as $\Delta^{-1} 
\simeq 6\%$ for $m_{\tilde{t}} \simeq 800~{\rm GeV}$ (and $M_{\rm Higgs} 
\simeq 114.4~{\rm GeV}$).  The fine-tuning becomes even worse, $\Delta^{-1} 
\simeq 4\%$, if we use $m_{\tilde{t}} \simeq 1~{\rm TeV}$.

There are two directions one can consider.  One is to make 
$M_{\rm mess}$ really close to $m_{\tilde{t}}$.  Such a situation 
may arise if we introduce non-renormalizable physics at the TeV scale 
(e.g.~\cite{Barbieri:2002sw}), but then we lose the supersymmetric 
desert and physics associated with it, such as gauge coupling 
unification.  The reduction of fine-tuning is also mild, and a 
cancellation of order $10\%$ is still required.  The other is to 
introduce a rather large stop mixing parameter $A_t$, which has 
implicitly been assumed to be small in the above analysis.  This 
allows $M_{\rm Higgs} \simgt 114.4~{\rm GeV}$ even with $m_{\tilde{t}}$ 
as small as $\simeq (400\!\sim\!500)~{\rm GeV}$.  We must, however, 
then generate such a large $A_t$, keeping $M_{\rm mess}$ small and 
without introducing the supersymmetric flavor problem.  The implication 
for the fine-tuning is also not obvious, because the large $A_t$ 
gives an additional negative contribution to $m_{H_u}^2$ so that 
the entire problem should be reanalyzed.

Recently, it has been observed~\cite{Choi:2005uz} that the mediation 
scale of supersymmetry breaking, $M_{\rm mess}$, can be effectively 
lowered in a scenario in which the moduli~\cite{Kaplunovsky:1993rd} 
and anomaly mediated~\cite{Randall:1998uk,Giudice:1998xp} contributions 
to supersymmetry breaking are comparable, which occurs naturally in a 
low energy limit of certain string-motivated setup~[\ref{Choi:2004sx:X}~%
--~\ref{Kachru:2003aw:X}].  The point is that, while $M_{\rm mess}$ can 
be effectively lowered to the TeV region, there is no physical threshold 
associated with it; the lowering occurs because of an interplay between 
the moduli and anomaly mediated contributions in the renormalization 
group evolutions.%
\footnote{For alternative possibilities for reducing $M_{\rm mess}$, 
or canceling the top-stop contribution, see e.g.~[\ref{Feng:1999mn:X}~%
--~\ref{Birkedal:2004xi:X}].}
A remarkable consequence of such a low effective messenger scale is 
that the superparticle masses can be unified at the TeV scale.  An 
interesting recent suggestion is that this can be used to address the 
issue of fine-tuning by making $M_{\rm mess}$ close to $m_{\tilde{t}}$ 
and thus by producing a little hierarchy between the Higgs boson and 
the other scalar squared masses~\cite{Choi:2005hd}.  This, however, 
can hardly be the only essence for the solution to the supersymmetric 
fine-tuning problem, which we define such that the amount of cancellation 
can be made smaller than 1 in 5.  Since the bound on $m_{\tilde{t}}$ is 
still strong even for the unit logarithm, $\ln(M_{\rm mess}/m_{\tilde{t}}) 
= 1$, it is not clear that the tight tension with $M_{\rm Higgs}$ really 
allows $\Delta^{-1} \geq 20\%$. 

In this paper we show that it is possible to obtain $\Delta^{-1} \geq 
20\%$ along this direction.  We find that a crucial element for the 
reduction of fine-tuning is, besides a very small $M_{\rm mess}$, 
a rather large value for $A_t$, which necessarily arises as a consequence 
of lowering the effective messenger scale, $M_{\rm mess}$.  The result is 
not trivial because the large $A_t$ gives a sizable contribution to the 
right-hand-side of Eq.~(\ref{eq:corr-Higgs}), giving a stronger bound 
on naturalness than the case with small $A_t$.  In fact, we obtain $|\delta 
m_{H_u, {\rm top}}^2|^{1/2}/m_{\tilde{t}} \simeq 0.4 \gg (1/8\pi^2)^{1/2}$ 
even for the unit logarithm, so that there is no real hierarchy between 
$m_{\tilde{t}}$ and the Higgs soft mass.  We find that $\Delta^{-1} \geq 
20\%$ can be obtained only if the superparticle masses are in certain 
restricted ranges.  This in turn provides a set of rather sharp predictions 
on the superparticle masses.  A somewhat surprising result is that one 
of the top squarks can be rather light; even $m_{\tilde{t}_1} \simlt 
300~{\rm GeV}$ is allowed due to large $A_t$.  This may be relevant for 
Run~II at the Tevatron.

In the simplest setup, we obtain the following predictions on the 
spectrum.  The gauginos are almost universal at the weak scale, as 
well as the squarks and sleptons:
\begin{equation}
  m_{\tilde{b}} \simeq m_{\tilde{w}} \simeq m_{\tilde{g}} 
    \simeq M_0,
\label{eq:gaugino}
\end{equation}
\begin{equation}
  m_{\tilde{q}} \simeq m_{\tilde{u}} \simeq m_{\tilde{d}} 
    \simeq m_{\tilde{l}} \simeq m_{\tilde{e}} 
    \simeq \frac{M_0}{\sqrt{2}},
\label{eq:sfermion}
\end{equation}
where the parameter $M_0$ is in the range
\begin{equation}
  450~{\rm GeV} \simlt M_0 \simlt 900~{\rm GeV}.
\label{eq:M0-range}
\end{equation}
The upper bound on $M_0$ arises from requiring $\Delta^{-1} \geq 20\%$. 
The top squark masses have appreciable splittings from $M_0/\sqrt{2}$:
\begin{equation}
  m_{\tilde{t}_{1,2}} \simeq 
    \frac{M_0 \mp m_t}{\sqrt{2}}.
\label{eq:stop}
\end{equation}
The lightest Higgs boson mass is bound by
\begin{equation}
  M_{\rm Higgs} \simlt 120~{\rm GeV},
\label{eq:Higgs}
\end{equation}
and the ratio of the vacuum expectation values (VEVs) for the two Higgs 
doublets, $\tan\beta \equiv \langle H_u \rangle/\langle H_d \rangle$, 
satisfies
\begin{equation}
  \tan\beta \simgt 5.
\label{eq:tan-beta}
\end{equation}
The lightest superparticle is the neutral Higgsino of the mass
\begin{equation}
  m_{\tilde{h}^0} \simlt 190~{\rm GeV},
\label{eq:Higgsinos}
\end{equation}
which is nearly degenerate with the charged Higgsino:
\begin{equation}
  m_{\tilde{h}^{\pm}} - m_{\tilde{h}^0} 
    = \frac{m_Z^2}{2 M_0} (1 + \epsilon),
\label{eq:H-split}
\end{equation}
where $|\epsilon| \simlt 0.2$ in the relevant parameter region. 
The bounds in Eqs.~(\ref{eq:Higgs},~\ref{eq:Higgsinos}) follow from 
$\Delta^{-1} \geq 20\%$.  The upper bounds in Eq.~(\ref{eq:M0-range}) 
and Eq.~(\ref{eq:Higgsinos}) can be relaxed to $1300~{\rm GeV}$ and 
$270~{\rm GeV}$, respectively, if we allow $\Delta^{-1}$ as small 
as $10\%$.  The bound in Eq.~(\ref{eq:tan-beta}) is also relaxed 
to $\tan\beta \simgt 4$.

In the next section we describe the framework.  The analysis of 
fine-tuning is given in section~\ref{sec:ewsb}, together with the 
resulting predictions on the spectrum.  The importance of a small 
$B$ parameter is pointed out, and a possible mechanism to realize 
it is presented.  Cosmological implications are also discussed.

\section{Framework}
\label{eq:framework}

Following Refs.~\cite{Choi:2004sx,Endo:2005uy}, we consider that the 
dominant source of supersymmetry breaking in the MSSM sector comes from 
the $F$-term VEVs of the chiral compensator superfield $C$ and a single 
moduli field $T$.  The effective supergravity action is given by
\begin{eqnarray}
  S &=& \int\! d^4 x\, \sqrt{-g}\, \Biggl[ \int\! d^4\theta\, 
    C^\dagger C {\cal F} - \theta^2 \bar{\theta}^2 
    C^{\dagger 2} C^2 {\cal P}_{\rm lift}
\nonumber\\
  && {} + \Biggl\{ \int\! d^2\theta\, \Bigl( \frac{1}{4} f_a 
    {\cal W}^{a \alpha} {\cal W}^a_\alpha + C^3 W \Bigr) + {\rm h.c.}
    \Biggr\} \Biggl],
\label{eq:action}
\end{eqnarray}
where $g_{\mu\nu}$ is the metric in the superconformal frame.  The 
function ${\cal F}$, which is related to the K\"ahler potential $K$ 
by ${\cal F} = -3 \exp(-K/3)$, the superpotential $W$, and the gauge 
kinetic functions $f_a$ take the form
\begin{eqnarray}
  {\cal F} &=& {\cal F}_0(T+T^\dagger) 
    + (T+T^\dagger)^{r_i} \Phi_i^\dagger \Phi_i,
\label{eq:F} \\
  W &=& W_0(T) + W_{\rm Yukawa},
\label{eq:W} \\
  f_a &=& T,
\label{eq:fa}
\end{eqnarray}
where $W_{\rm Yukawa}$ is the MSSM Yukawa couplings, and we have assumed 
that the nonlinear transformation acting on ${\rm Im} T$ is (approximately) 
preserved in ${\cal F}$.  The second term in Eq.~(\ref{eq:action}) is 
introduced to allow the cancellation of the cosmological constant at 
the minimum of the potential.  This term is supposed to take the form
\begin{equation}
  {\cal P}_{\rm lift} = d(T+T^\dagger)^{n_P},
\label{eq:Plift}
\end{equation}
which should arise from supersymmetry breaking in some sector that 
does not give direct contributions to supersymmetry breaking in the 
MSSM sector.  The functions ${\cal F}_0$ and $W_0$ are taken to be
\begin{eqnarray}
  {\cal F}_0(T+T^\dagger) &=& -3 (T+T^\dagger)^{n_0/3} + \cdots,
\label{eq:F0}
\\
  W_0(T) &=& w_0 - A e^{-a T},
\label{eq:W0}
\end{eqnarray}
where the dots in ${\cal F}_0$ represent possible higher order threshold 
corrections at the compactification and/or string scales.  The parameters 
$w_0$ and $A$ are assumed to be of order $m_{3/2} M_*^2 \ll M_*^3$ and 
$M_*^3$, respectively, where $m_{3/2}$ is the gravitino mass and $M_*$ 
the fundamental scale of order the Planck scale.%
\footnote{Realizing $w_0 \ll M_*^3$ in the context of particular string 
theory may require fine-tuning~\cite{Choi:2004sx}.}
These parameters can be taken real under the presence of an approximate 
shift symmetry for ${\rm Im} T$.  The parameter $a$ is a real constant 
of order $8\pi^2$: $a=8\pi^2/N$, where $N$ is the size of the gauge 
group generating the nonperturbative superpotential for $T$ through 
gaugino condensation.  An interesting candidate for $T$ is the 
moduli parameterizing the volume of the $n$ compact extra dimensions: 
$T \propto R^n$, where $R$ is the length scale for the compact space. 
In this case $n_0 = 3$, and $r_i = \tilde{n}_i/n$ if $\Phi_i$ propagates 
in $(4+\tilde{n}_i)$-dimensional subspace.

The superpotential of Eq.~(\ref{eq:W0}) stabilizes the modulus $T$ and 
produces the $F$-term VEVs for $C$ and $T$.  At the leading order in 
$1/\ln(A/w_0) \sim 1/\ln(M_{\rm Pl}/m_{3/2})$, 
\begin{eqnarray}
  a T &=& \ln\biggl(\frac{M_{\rm Pl}}{m_{3/2}}\biggr),
\label{eq:aT} \\
  \frac{m_{3/2}}{M_0} &=& \frac{2}{3} 
    \frac{\partial_T K_0}{\partial_T \ln(V_{\rm lift})} 
    \ln\biggl(\frac{M_{\rm Pl}}{m_{3/2}}\biggr),
\label{eq:ratio}
\end{eqnarray}
where $K_0 = -3 \ln(-{\cal F}_0/3)$, $M_{\rm Pl}$ is the reduced Planck 
scale, $m_{3/2} = e^{K_0/2} W_0$ is the gravitino mass, which is taken 
to be $m_{3/2} \approx (10\!\sim\!100)~{\rm TeV}$, and $V_{\rm lift} 
= e^{2 K_0/3} {\cal P}_{\rm lift}$ is the up-lifting potential.  The 
quantity $M_0$ is defined by
\begin{equation}
  M_0 = \frac{F_T}{T+T^\dagger}.
\label{eq:M0}
\end{equation}
To obtain the simple relation of Eq.~(\ref{eq:ratio}), it is essential 
that $T$ is stabilized by a single exponential factor in $W_0$.

An interesting property of this setup is that it allows a reduction 
of the effective messenger scale, $M_{\rm mess}$, due to an interplay 
between the moduli and anomaly mediated contributions~\cite{Choi:2005uz}. 
Suppose that only the appreciable Yukawa couplings, $W = (\lambda_{ijk}/6) 
\Phi_i \Phi_j \Phi_k$ with $y_{ijk} \equiv \lambda_{ijk}/(Z_i Z_j 
Z_k)^{1/2} \simgt \sqrt{8\pi^2}$, are among the fields satisfying 
$r_i + r_j + r_k = 1$, where $Z_i$'s are the wavefunction renormalization 
factors, $Z_i = (T+T^\dagger)^{r_i}$.  Suppose also that $r_i$'s 
satisfy $\sum_i r_i Y_i = 0$, where $Y_i$ is the hypercharge of $\Phi_i$. 
In this case the low-energy soft supersymmetry breaking parameters, 
defined by
\begin{equation}
  {\cal L}_{\rm soft} 
    = -\frac{1}{2} M_a \lambda^a \lambda^a - m_i^2 |\phi_i|^2 
      - \frac{1}{6}(A_{ijk} y_{ijk} \phi_i \phi_i \phi_k + {\rm h.c.}),
\label{eq:Lsoft}
\end{equation}
are given by
\begin{eqnarray}
  M_a(\mu_R) &=& M_0 \Biggl[ 1 - \frac{b_a}{8\pi^2} g_a^2(\mu_R) 
    \ln\Biggl( \frac{M_{\rm mess}}{\mu_R} \Biggr) \Biggr],
\label{eq:Ma-LE} \\
  A_{ijk}(\mu_R) &=& M_0 \Biggl[ (r_i+r_j+r_k) 
    - 2 \Bigl\{ \gamma_i(\mu_R)+\gamma_j(\mu_R)+\gamma_k(\mu_R) \Bigr\} 
    \ln\Biggl( \frac{M_{\rm mess}}{\mu_R} \Biggr) \Biggr],
\label{eq:Aijk-LE} \\
  m_i^2(\mu_R) &=& M_0^2 \Biggl[ r_i - 4\, \Biggl\{ 
    \gamma_i(\mu_R) - \frac{1}{2} \frac{d\gamma_i(\mu_R)}{d \ln\mu_R} 
    \ln\Biggl( \frac{M_{\rm mess}}{\mu_R} \Biggr) \Biggr\} 
    \ln\Biggl( \frac{M_{\rm mess}}{\mu_R} \Biggr) \Biggr],
\label{eq:mi2-LE}
\end{eqnarray}
including one-loop renormalization group effects.  Here, $g_a^2(\mu_R)$ 
are the running gauge couplings at a scale $\mu_R$, $b_a$ and 
$\gamma(\mu_R)$ are the beta-function coefficients and the anomalous 
dimensions defined by $d(1/g_a^2)/d\ln\mu_R=-b_a/8\pi^2$ and 
$d \ln Z_i/d \ln\mu_R = -2 \gamma_i$, respectively.  The parameter 
$M_{\rm mess}$ is given by
\begin{equation}
  M_{\rm mess} = \frac{M_{\rm GUT}}{(M_{\rm Pl}/m_{3/2})^{\alpha/2}},
\label{eq:eff-mess}
\end{equation}
where $M_{\rm GUT}$ is the scale at which the effective action of 
Eq.~(\ref{eq:action}) is given, which is supposed to be the unification 
(compactification) scale, and $\alpha$ parameterizes the ratio of the 
anomaly-mediated to the moduli contributions:
\begin{equation}
  \alpha \equiv \frac{m_{3/2}}{M_0 \ln(M_{\rm Pl}/m_{3/2})}
    = \frac{2 n_0}{2 n_0 - 3 n_P} + \cdots.
\label{eq:alpha}
\end{equation}
The last equality of Eq.~(\ref{eq:alpha}) follows from Eq.~(\ref{eq:ratio}) 
with Eqs.~(\ref{eq:Plift},~\ref{eq:F0}).  Equations~(\ref{eq:Ma-LE}~--~%
\ref{eq:mi2-LE}) show that the low energy values for the soft supersymmetry 
breaking parameters are obtained simply by setting the moduli-dominated 
supersymmetry breaking boundary conditions at the scale $M_{\rm mess}$ and 
then evolving them to the scale $\mu_R$.  In this sense, $M_{\rm mess}$ 
effectively plays a role of the messenger scale. 

The idea suggested in Ref.~\cite{Choi:2005hd} is to use this feature 
to make a little hierarchy between the Higgs boson and the other 
scalar squared masses.  For this purpose, $M_{\rm mess}$ must be 
close to the TeV scale, requiring $\alpha \simeq 2$ to a high degree. 
A simple way to achieve this is to choose $n_0 = 3$ and $n_P = 1$ 
(see Eq.~(\ref{eq:alpha})).  This is, however, not entirely enough; 
we also have to check that higher order terms of ${\cal F}_0$ in 
Eq.~(\ref{eq:F0}), given by powers of $1/(T+T^\dagger)$, do not give 
significant corrections to $\alpha$.  Since $\langle T + T^\dagger 
\rangle \simeq 2/g_{\rm GUT}^2$ is not so large, this requires the 
coefficients of these terms to be somewhat suppressed.  From the 
effective field theory point of view, this is technically natural 
because nothing is strongly coupled below the scale $\approx M_{\rm GUT}$. 
In the string theory context, this may require both the compactification 
volume and the string coupling to be taken somewhat large. In any case, 
having seen that $\alpha \simeq 2$ can be obtained without fine-tuning 
parameters, we simply treat $M_{\rm mess}$ as a free parameter of 
order the TeV scale.  The sensitivity of the weak scale to $\alpha$, 
or $M_{\rm mess}$, will be included in our analysis of the supersymmetric 
fine-tuning problem.

The moduli couplings to the matter and Higgs fields, $r_i$, must 
also be chosen such that the renormalization group properties of 
Eqs.~(\ref{eq:Ma-LE}~--~\ref{eq:mi2-LE}) are preserved.  Here we impose 
the following conditions to determine the values of $r_i$.  (i) Motivated 
by the successful $b/\tau$ Yukawa unification, we assume that the third 
generation matter is embedded into $SU(5)$ representations.  (ii) To 
avoid the supersymmetric flavor problem, $r_i$ must be assigned flavor 
universal.  These two requirements lead to the following assignment 
for $r_i$:
\begin{equation}
  r_{Q} = r_{U} = r_{E} = \frac{1}{2}, \qquad
  r_{D} = r_{L}, \qquad
  r_{H_u} = r_{H_d} = 0.
\label{eq:model-su5}
\end{equation}
Note that the $SU(5)$ requirement of (i), together with $\sum_i r_i 
Y_i = 0$, necessarily leads to $r_{H_u} = r_{H_d}$.  In the simplest 
case that all the matter fields have a common $r_i$ (propagate in the 
same spacetime dimensions), we obtain 
\begin{equation}
  r_{Q} = r_{U} = r_{D} = r_{L} = r_{E} = \frac{1}{2}, \qquad
  r_{H_u} = r_{H_d} = 0.
\label{eq:model-univ}
\end{equation}
This can be obtained in the string theory construction if the matter 
fields live on intersections of $D7$ branes while the Higgs fields 
on $D3$ branes.%
\footnote{In the string theory realization based on~\cite{Kachru:2003aw}, 
the modulus $T$ is related to the compactification radius $R$ by 
$T \propto R^4$.  Thus, the coefficient of $\Phi_i^\dagger \Phi_i$ in 
Eq.~(\ref{eq:F}) is proportional to $R^4$ ($R^0$) if $\Phi_i$ propagates 
on a $D7$ ($D3$) brane: $r_i = 1$ ($r_i = 0$).  If $\Phi_i$ lives on 
an intersection of $D7$ branes, the coefficient is $\propto R^2$: 
$r_i = 1/2$.}
Alternatively, it may arise in a six dimensional theory if matter 
and Higgs fields propagate in five and four dimensions, respectively 
($n=2$, $\tilde{n}_{\rm matter} = 1$ and $\tilde{n}_{\rm Higgs} = 0$). 
It is interesting that the choice of $r_i$ required by low energy 
phenomenology, Eq.~(\ref{eq:model-univ}), is so simple that it can 
be accommodated into the string theory (or extra dimensional) setup 
without much difficulties.  In our analysis for the fine-tuning, we 
only consider the universal matter model of Eq.~(\ref{eq:model-univ}). 
More general cases of Eq.~(\ref{eq:model-su5}), however, could also 
work if $\tan\beta$ is small, e.g. $\tan\beta \simlt 10$, so that the 
renormalization group contribution to $m_{H_d}^2$ from the bottom Yukawa 
coupling is sufficiently small.

The assignment of Eq.~(\ref{eq:model-univ}) leads to the following flavor 
universal soft supersymmetry breaking parameters
\begin{equation}
  M_1 = M_2 = M_3 = M_0,
\label{eq:gaugino-masses}
\end{equation}
\begin{equation}
  m_{\tilde{q}}^2 = m_{\tilde{u}}^2 = m_{\tilde{d}}^2 
    = m_{\tilde{l}}^2 = m_{\tilde{e}}^2 = \frac{M_0^2}{2},
\label{eq:sfermion-masses}
\end{equation}
\begin{equation}
  A_u = A_d = A_e = M_0,
\label{eq:Aterms}
\end{equation}
at the effective messenger scale $M_{\rm mess}$, which is very close 
to the TeV scale (see the analysis in the next section).  Here, $A_u$, 
$A_d$, and $A_e$ represent the trilinear scalar couplings for the 
up-type squarks, down-type squarks and charged sleptons, respectively. 
As we will see in the next section, a particularly important feature 
of Eqs.~(\ref{eq:gaugino-masses}~--~\ref{eq:Aterms}) in solving the 
supersymmetric fine-tuning problem is rather large values for the $A$ 
parameters, specifically that for the top squark, $A_t$, which necessarily 
arise as a consequence of lowering $M_{\rm mess}$.  This allows us 
to evade the LEP~II bound on the Higgs boson mass with relatively small 
top squark masses.  The situation, however, is not trivial because the 
large $A_t$ also gives an additional negative radiative correction to 
$m_{H_u}^2$.  We thus have to study the dynamics of electroweak symmetry 
breaking more carefully, to see that the supersymmetric fine-tuning 
problem can actually be solved.

\section{Electroweak Symmetry Breaking without Fine-Tuning}
\label{sec:ewsb}

As discussed in the introduction, the amount of fine-tuning is roughly 
measured by the ratio of the left-hand-side to the largest contribution 
to the right-hand-side of the equation determining the weak scale, 
$M_{\rm Higgs}^2/2 \simeq -m_{H_u}^2 - |\mu|^2$.  For the model of 
Eq.~(\ref{eq:model-univ}), the correction to the Higgs mass-squared 
parameter is given by
\begin{equation}
  \delta m_{H_u}^2 \simeq 
    - \frac{3 y_t^2}{4\pi^2} 
      M_0^2 \ln\Biggl( \frac{M_{\rm mess}}{m_{\tilde{t}}} \Biggr)
    + \frac{15 g_2^2 + 3 g_1^2}{40\pi^2} 
      M_0^2 \ln\Biggl( \frac{M_{\rm mess}}{m_{\lambda}} \Biggr),
\label{eq:delta}
\end{equation}
where the first and second terms represent contributions from the top 
Yukawa coupling and the gauge couplings, respectively, and $m_{\tilde{t}} 
\simeq M_0/\sqrt{2}$ and $m_{\lambda} \simeq M_0$.  Here, the first 
term includes the effects from both $m_{\tilde{t}}$ and $A_t$.  The 
fine-tuning is then given by the ratio of $M_{\rm Higgs}^2/2$ to the 
largest of the two terms in Eq.~(\ref{eq:delta}). 

One might naively think that the fine-tuning from the top Yukawa 
contribution may be entirely eliminated by choosing $M_{\rm mess}$ very 
close to the value with which the logarithm of the first term almost 
vanishes.  This, however, leads to another fine-tuning of the parameter 
$M_{\rm mess}$.  In general, the amount of fine-tuning for the parameter 
$a_i$ can be obtained by studying the logarithmic sensitivity of $v^2$ 
to $a_i$ using appropriate measures~\cite{Chacko:2005ra-2}.  In our 
context, this leads to the following definition for the fine-tuning 
parameter
\begin{equation}
  \Delta^{-1} = \frac{M_{\rm Higgs}^2}{2} {\Bigg /} 
    {\rm max} \Biggl\{\, \Biggl| \frac{3 y_t^2}{4\pi^2} 
      M_0^2 \ln\biggl( \frac{M_{\rm mess}}{m_{\tilde{t}}} \biggr) 
      \Biggr|,\,\, \Biggl| \biggl( \frac{3 y_t^2}{4\pi^2} - 
      \frac{15 g_2^2 + 3 g_1^2}{40\pi^2} \biggr) M_0^2 \Biggr|\, \Biggr\}.
\label{eq:ft-param}
\end{equation}
The first and second factors in the denominator measure the sensitivity 
of the weak scale to the top contribution (the top Yukawa coupling) and 
the effective messenger scale (the ratio of the moduli to anomaly mediated 
contributions), respectively. 

Equation~(\ref{eq:ft-param}) tells us that the fine-tuning becomes 
better if we lower $M_0$, since the sensitivity of $M_{\rm Higgs}$ to 
$M_0$ is much weaker than that of the denominator.  There is, however, 
a lower bound on $M_0$, coming from the bound on the Higgs boson mass, 
$M_{\rm Higgs} \simgt 114.4~{\rm GeV}$.  The question is whether there 
is a parameter region in which $\Delta^{-1} \geq 20\%$, still satisfying \
$M_{\rm Higgs} \simgt 114.4~{\rm GeV}$.  

There are four parameters $\mu$, $B$, $m_{H_u}^2$ and $m_{H_d}^2$ in 
the Higgs potential which need to be fixed to calculate $M_{\rm Higgs}$ 
and $\Delta^{-1}$.  For our choice of $r_{H_u} = r_{H_d} = 0$, the 
Higgs soft mass-squared parameters $m_{H_u}^2$ and $m_{H_d}^2$ vanish 
at $M_{\rm mess}$, at the leading order in $1/aT \approx 1/8 \pi^2$ 
expansion.  There are, however, non-vanishing corrections arising at 
the next-to-leading order, which depend on unknown threshold effects 
at the scale $M_{\rm GUT}$.  The existence of higher order corrections 
also raises the question if similar corrections to the other scalars, 
specifically to the top squarks, give too large contributions to $\delta 
m_{H_u}^2$ through renormalization group evolution because there is 
no reason that the large logarithm, $\ln(M_{\rm GUT}/m_{\tilde{t}})$, 
disappears for such contributions.  In order to discuss fine-tuning 
based on Eq.~(\ref{eq:ft-param}), all these corrections to $m_{H_u}^2$, 
including the one from higher loop renormalization group evolution, 
must be smaller than the larger of the two quantities in the dominator 
of Eq.~(\ref{eq:ft-param}).  We find, however, that if the higher order 
corrections to the soft masses are smaller than about $v^2$, which is 
quite natural given that they are $O(M_0^2/8\pi^2)$ effects, all the 
unknown contributions to $m_{H_u}^2$ can be made sufficiently small. 
This implies that we can simply treat $m_{H_u}^2$ and $m_{H_d}^2$ as 
free parameters at the scale $M_{\rm mess}$ and analyze fine-tuning 
using Eq.~(\ref{eq:ft-param}).  The sizes of these parameters are 
bounded roughly as $m_{H_u}^2, m_{H_d}^2 \simlt (M_{\rm Higgs}^2/2)/20\% 
\approx (200~{\rm GeV})^2$ if we require $\Delta^{-1} \geq 20\%$. 
The unknown corrections also induce uncertainties for the predictions 
of the squark and slepton masses, but they are at most of order 
$v^2/M_0$ and thus small.%
\footnote{We assume that these higher order corrections are flavor 
universal, at least approximately, so that the supersymmetric flavor 
problem is not reintroduced.}

The natural sizes of $\mu$ and $B$ parameters in supergravity are 
$O(m_{3/2})$, which is unacceptably large.  This requires some 
mechanism of suppressing these parameters such as the ones proposed 
in Refs.~\cite{Randall:1998uk,Pomarol:1999ie}.  In fact, the sizes of 
$\mu$ and $B$ are severely constrained in our framework.  The $\mu$ 
parameter is subject to the bound $|\mu| \simlt 190~{\rm GeV}$, given 
by the naturalness requirement of $M_{\rm Higgs}^2/2|\mu|^2 \geq 20\%$. 
The value of $B$ must also be small.  Since both $|\mu|^2$ and 
$m_{H_d}^2$ are bounded by $\approx (200~{\rm GeV})^2$, the value 
of $B$ must be $\approx (350/\tan\beta)~{\rm GeV}$ or smaller. 
Since $\tan\beta$ must be larger than about $5$, as we will see 
later, this requires rather small values for the $B$ parameter.

Is it possible to obtain such a small $B$ without fine-tuning? 
Suppose there is a singlet field $\Sigma$ that is subject to an 
approximate shift symmetry and couples to the Higgs doublets as
\begin{equation}
  \delta S = \int\! d^4 x\, \sqrt{-g}\, 
    \int\! d^4\theta\, C^\dagger C \Bigl( \kappa 
    (\Sigma + \Sigma^\dagger) H_u H_d + {\rm h.c.} \Bigr),
\label{eq:mu-gen}
\end{equation}
where $\kappa$ is a constant and we have normalized $\Sigma$ as 
a dimensionless field.  Suppose also that the $\Sigma$ field has 
a vanishing VEV in the lowest component, but has a non-vanishing VEV 
of order $M_0$ in the $F$ component that is real in the basis where 
$F_C$ is real: $\langle \Sigma \rangle = \theta^2 F_\Sigma$. 
In this case the interaction of Eq.~(\ref{eq:mu-gen}) gives 
\begin{eqnarray}
  \mu &=& \kappa F_\Sigma,
\label{eq:mu-GUT} \\
  B &=& (\gamma_{H_u}+\gamma_{H_d}) m_{3/2},
\label{eq:B-GUT}
\end{eqnarray}
at the scale $\mu_R \simeq M_{\rm GUT}$, so that $\mu$ and 
$B$ of order $F_\Sigma \sim M_0 \ll m_{3/2}$ are naturally 
obtained~\cite{Randall:1998uk}.%
\footnote{The $\Sigma$ field should not have a coupling to the MSSM 
fields other than that in Eq.~(\ref{eq:mu-gen}), at the lowest order. 
This can be easily achieved in higher dimensional theories.}
Note that the contribution from the moduli $F$-term, $F_T$, is 
absent in Eq.~(\ref{eq:B-GUT}) with our choice of $r_{H_u} = r_{H_d} 
= 0$.  This size of $B$, however, is still too large, since we 
need $B \approx (10\!\sim\!70)~{\rm GeV}$ while $M_0$ is of order 
$500~{\rm GeV}\!\sim\!1~{\rm TeV}$, as we will see later. 

A further suppression of $B$, however, arises through renormalization 
group evolution.  Given the $B$ parameter of Eq.~(\ref{eq:B-GUT}) at 
$\mu_R \simeq M_{\rm GUT}$, one-loop renormalization group evolution 
of $B$ gives 
\begin{equation}
  B(\mu_R) = -2 M_0 \Bigl\{ \gamma_{H_u}(\mu_R)+\gamma_{H_d}(\mu_R) 
    \Bigr\} \ln\Biggl( \frac{M_{\rm mess}}{\mu_R} \Biggr),
\label{eq:B-LE}
\end{equation}
at lower energies, where we have used Eqs.~(\ref{eq:Ma-LE}~--%
~\ref{eq:mi2-LE}) for the other soft masses.  We find that the $B$ 
parameter vanishes at the scale $M_{\rm mess}$, which is close to TeV. 
Since there is no large correction to the $\mu$ parameter, we can naturally 
obtain the desired values of $\mu \approx (100\!\sim\!200)~{\rm GeV}$ and 
$B \approx (10\!\sim\!70)~{\rm GeV}$ with an $O(1)$ value for $\kappa$. 
In fact, this property is quite general.  Any mechanism that eliminates 
the classical contribution to $B$ of order $m_{3/2}$ and leaves the 
``anomaly mediated'' contribution of Eq.~(\ref{eq:B-GUT}) leads to 
$B \approx 0$ at $\mu_R \simeq M_{\rm mess}$.  This is important 
to eliminate fine-tuning entirely from electroweak symmetry breaking. 
Having seen Eqs.~(\ref{eq:mu-GUT}~--~\ref{eq:B-LE}), we treat $\mu$ 
and $B$ as free parameters in our analysis below.%
\footnote{The explicit mechanism discussed here requires $F_\Sigma$ 
to be (almost) real in the basis where $F_C$ is real.  Then, if the 
weak scale value of $B$ is given by Eq.~(\ref{eq:B-LE}) without an 
additional contribution, there is no supersymmetric $CP$ problem. 
It is interesting that the value of $B$ in Eq.~(\ref{eq:B-LE}) is, 
in fact, consistent with large values of $\tan\beta$, e.g. $\tan\beta 
\simgt 10$, for $M_{\rm mess}$ about a TeV.}

\begin{figure}[t]
\begin{center}
  \includegraphics[height=7.5cm]{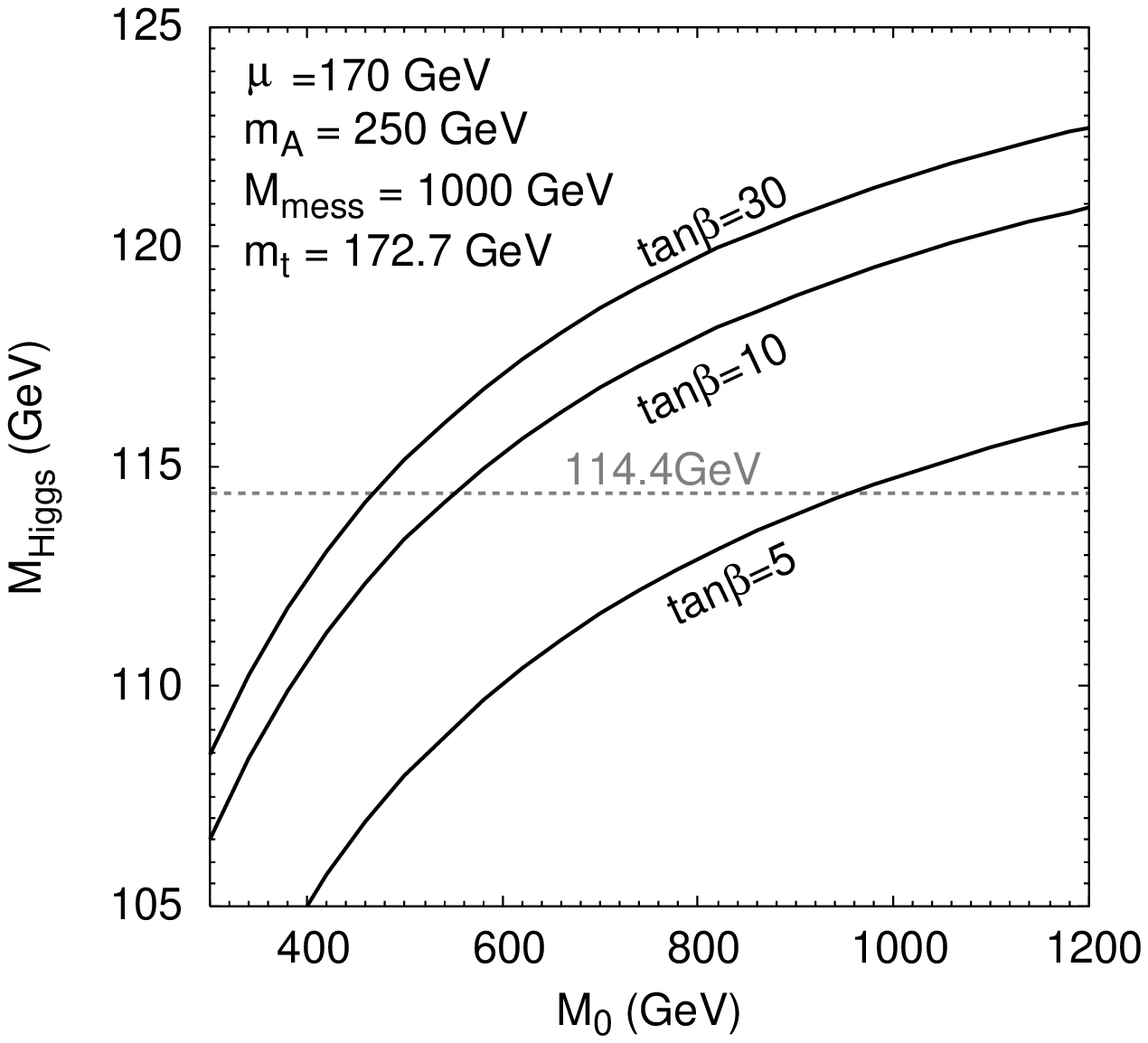}
  \includegraphics[height=7.5cm]{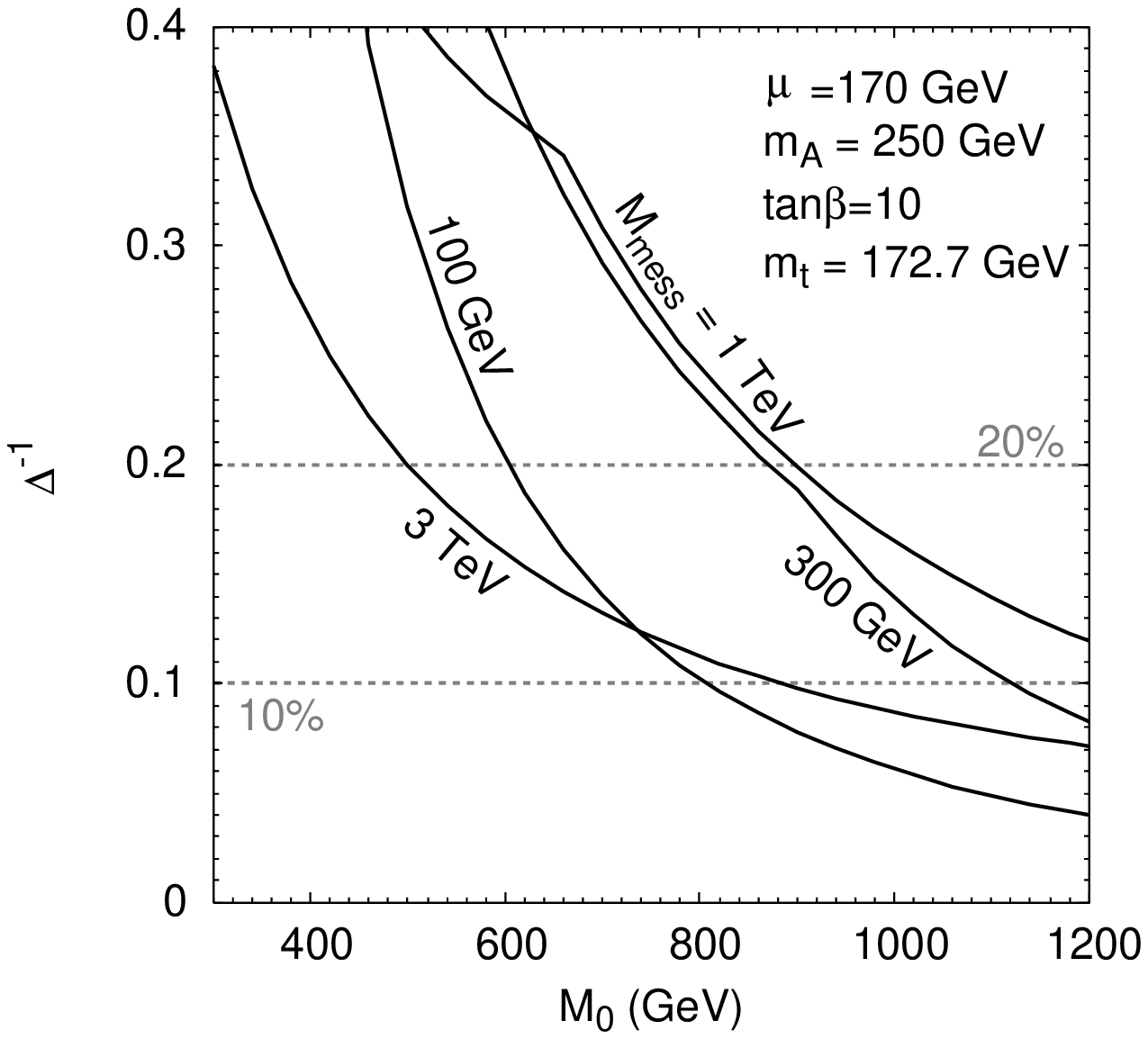}
\end{center}
\caption{The Higgs boson mass, $M_{\rm Higgs}$, and the fine-tuning 
 parameter, $\Delta^{-1}$, as a function of $M_0$.  The lower bound on 
 $M_0$ is obtained from the left figure using the experimental bound 
 of $M_{\rm Higgs} \simgt 114.4~{\rm GeV}$, while the upper bound can 
 be read off from the right figure with the requirement of $\Delta^{-1} 
 \geq 20\%$.}
\label{fig:m0-region}
\end{figure}

Now we analyze if we have a region of $M_0$ satisfying the experimental 
bound $M_{\rm Higgs} \simgt 114.4~{\rm GeV}$ and $\Delta^{-1} \geq 20\%$ 
simultaneously.  Out of four free parameters $m_{H_u}^2$, $m_{H_d}^2$, 
$\mu$ and $B$, one combination is fixed by the vacuum expectation value 
$v$, leaving three independent parameters, which we take to be $\mu$, 
$m_A$ and $\tan\beta$, where $m_A$ is the mass of the pseudo-scalar 
Higgs boson.  The summary of the analysis is presented in 
Fig.~\ref{fig:m0-region}.
In the left figure, we show the Higgs boson mass $M_{\rm Higgs}$, 
calculated using {\it FeynHiggs~2.2}~\cite{Heinemeyer:1998yj}, as a function 
of $M_0$ for various values of $\tan\beta$.  The other parameters are 
fixed as $\mu = 170~{\rm GeV}$, $m_A = 250~{\rm GeV}$ and $M_{\rm mess} 
= 1~{\rm TeV}$.  The sensitivity of $M_{\rm mess}$ to these parameters 
are rather mild.  For the top quark mass, we have used the latest 
central value of $m_t = 172.7~{\rm GeV}$~\cite{Group:2005cc}.  From 
the figure, we obtain the lower bound on $M_0$, depending on the value 
of $\tan\beta$.  For $\tan\beta = 10$, for example, the bound is $M_0 
\simgt 550~{\rm GeV}$.  The bound becomes weaker for larger $\tan\beta$; 
it becomes $M_0 \simgt 450~{\rm GeV}$ for $\tan\beta = 30$.  The bound 
stays practically the same for $\tan\beta \simgt 30$. 

It is remarkable that $M_0$ as low as $(450\!\sim\!550)~{\rm GeV}$ 
is allowed, which corresponds to $m_{\tilde{t}} \simeq M_0/\sqrt{2} 
\approx (320\!\sim\!390)~{\rm GeV}$.  Such low values of $m_{\tilde{t}}$ 
are consistent with the bound on the Higgs boson mass because of a 
large $A_t$, which necessarily arise as a consequence of lowering 
$M_{\rm mess}$.  In fact, the masses of the two top squarks have 
appreciable splittings from $M_0/\sqrt{2}$ because of the large $A_t$: 
$m_{\tilde{t}_{1,2}} \simeq (M_0 \mp m_t)/\sqrt{2}$.  The lighter top 
squark can thus be as light as $(200\!\sim\!270)~{\rm GeV}$.  This 
may be relevant for Run~II at the Tevatron. 

The $M_0$ dependence of the fine-tuning parameter $\Delta^{-1}$ is 
shown in the right figure.  Here we have fixed $\tan\beta$ and plotted 
$\Delta^{-1}$ for various values of $M_{\rm mess}$.  The sensitivities 
to the fixed parameters are again small.  From the figure, we can 
obtain the upper bound on $M_0$.  For a fixed value of $\Delta^{-1}$, 
the maximum value of $M_0$ is obtained for $M_{\rm mess} \approx 
(300\!\sim\!1000)~{\rm GeV}$; smaller values of $M_0$ are required 
both for larger and smaller $M_{\rm mess}$ because of the logarithmic 
enhancement of $m_{H_u}^2$ (the factor of $\ln(M_{\rm mess}/m_{\tilde{t}})$ 
in the denominator of Eq.~(\ref{eq:ft-param})).  Requiring the absence 
of fine-tuning, i.e. $\Delta^{-1} \geq 20\%$, the upper bound of $M_0 
\simlt 900~{\rm GeV}$ is obtained.  Compared with the lower bound 
previously derived, we find that there is a region of $M_0$ in which 
the bound on the Higgs boson mass and the requirement from naturalness 
are both satisfied.  For $\tan\beta = 30$ ($10$), it is $450~{\rm GeV} 
\simlt M_0 \simlt 900~{\rm GeV}$ ($550~{\rm GeV} \simlt M_0 \simlt 
900~{\rm GeV}$).  This shows that we have in fact solved the supersymmetric 
fine-tuning problem without extending the MSSM.  In order to have an 
allowed range for $M_0$, the effective messenger scale $M_{\rm mess}$ 
must be in the range $50~{\rm GeV} \simlt M_{\rm mess} \simlt 3~{\rm 
TeV}$, and $\tan \beta \simgt 5$ is necessary.  By requiring $M_0 \simlt 
900~{\rm GeV}$, we can also read off the upper bound on $M_{\rm Higgs}$ 
from the left figure, giving $M_{\rm Higgs} \simlt 120~{\rm GeV}$.

If we relax the naturalness criterion to $\Delta^{-1} \geq 10\%$, $M_0$ 
can be as large as $1300~{\rm GeV}$.  A larger range of $M_{\rm mess}$ 
is also allowed: $50~{\rm GeV} \simlt M_{\rm mess} \simlt 100~{\rm TeV}$. 
Here, the lower bound on $M_{\rm mess}$ comes from the fact that some 
of the superparticles become too light if $M_{\rm mess}$ is lowered 
beyond this value.  The bound on $\tan\beta$ is relaxed to $\tan\beta 
\simgt 4$.

The dependence of our results on the value of the top quark mass is 
not strong.  This is because the two loop effects on the Higgs boson 
mass, which give a sizable negative correction to $M_{\rm Higgs}$ for 
large superparticle masses, are not important here due to small values for 
$M_0$.  If we use $m_t = 178.0~{\rm GeV}$ instead of $172.7~{\rm GeV}$, for 
example, the lower bound on $M_0$ decreases from $\approx 450~{\rm GeV}$ 
to $\approx 400~{\rm GeV}$ and the upper bound on $M_{\rm Higgs}$ increases 
from $\simeq 120~{\rm GeV}$ to $\simeq 124~{\rm GeV}$, but there are no 
appreciable changes to the other numbers. 

The naturalness bound of $\mu \simlt 190~{\rm GeV}$ together with 
Eqs.~(\ref{eq:gaugino-masses}~--~\ref{eq:Aterms}) implies that the 
lightest supersymmetric particle is the Higgsino of mass about $|\mu|$. 
The mass splitting between the charged and neutral Higgsinos comes 
from the mixings with the gaugino states: $m_{\tilde{h}^{\pm}} - 
m_{\tilde{h}^0} = (m_Z^2/2 M_0)(1+\epsilon)$, where $|\epsilon| 
\simlt 0.2$ in the relevant parameter region, so that the neutral 
component is always lighter.  A similar bound of $m_{H_d}^2 \simlt 
(200~{\rm GeV})^2$ implies that the pseudo-scalar and charged Higgs 
bosons are relatively light: $m_A, m_{H^\pm} \simlt 300~{\rm GeV}$. 
This may have some implications on the rate of the rare $b \rightarrow 
s \gamma$ process, but the current theoretical status does not seem 
to allow us to make any definite statement~\cite{Neubert:2004dd}. 
The positive sign of $\mu$, however, seems to be preferred over the 
other one.

The range of the gravitino mass is determined as
\begin{equation}
  m_{3/2} \simeq 2 M_0 \ln\Biggl( \frac{M_{\rm Pl}}{m_{3/2}} \Biggr)
    \simeq (30\!\sim\!60)~{\rm TeV}.
\label{eq:gravitino}
\end{equation}
For $\Delta^{-1} \geq 10\%$, the upper bound can be relaxed to 
$80~{\rm TeV}$.  These values for $m_{3/2}$ are large enough to evade 
the cosmological gravitino problem.  The mass of the moduli field is 
a factor of $\ln(M_{\rm Pl}/m_{3/2})$ larger than the gravitino mass, 
$m_T \approx 1000~{\rm TeV}$, so that the cosmological moduli problem 
is also absent~\cite{Choi:2004sx,Endo:2005uy}.  The moduli field 
decays mainly into the gauge bosons, reheating the universe to about 
$100~{\rm MeV}$.  A small fraction of moduli, however, also goes 
into the gravitino, which in turn produces the neutral Higgsino 
through its decay.  This can provide dark matter of the universe, 
$\Omega_{\tilde{h}^0} \simeq 0.2$, for $m_{3/2}$ and $m_T$ in the 
range considered here~\cite{Kohri:2005ru}.  (The amount of the 
Higgsino produced by the direct moduli decay is negligible.)  This 
offers significant potential for the discovery at on-going dark 
matter detection experiments such as CDMS~II.

\section*{Acknowledgments}

The work of R.K. was supported by the U.S. Department of Energy under 
contract number DE-AC02-76SF00515.  The work of Y.N. was supported 
in part by the Director, Office of Science, Office of High Energy 
and Nuclear Physics, of the US Department of Energy under Contract 
DE-AC02-05CH11231, by the National Science Foundation under grant 
PHY-0403380, by a DOE Outstanding Junior Investigator award, and 
by an Alfred P. Sloan Research Fellowship.

\end{document}